\documentstyle[aps,subfigure,epsfig,amsmath,amssymb]{revtex} 
\begin{document} 
\newcommand {\cs}{$\clubsuit$}
\draft

\title{Demixing in mesoscopic boson-fermion clouds inside cylindrical
harmonic traps: quantum phase diagram and 
role of temperature}

\author{Z. Akdeniz$^{1,2}$, 
A. Minguzzi$^1$, 
P. Vignolo$^1$, and M.~P. Tosi$^1$}
\address{(1) NEST-INFM and Scuola Normale Superiore,
Piazza dei Cavalieri 7, I-56126 Pisa, Italy\\
(2) Department of Physics, University of Istanbul, Istanbul, Turkey}
\maketitle
\begin{abstract}
We use a semiclassical three-fluid thermodynamic model to evaluate the
phenomena of spatial demixing in mesoscopic clouds of fermionic and
bosonic atoms at high dilution under harmonic confinement, assuming
repulsive boson-boson and boson-fermion interactions and including
account of a bosonic thermal cloud at finite temperature $T$. The
finite system size allows three different regimes for the equilibrium density
profiles at $T=0$: a fully mixed state, a partially mixed state in
which the overlap between the boson and fermion clouds is decreasing,
and a fully demixed state where the two clouds have zero overlap. We
propose simple analytical rules for the two cross-overs between the
three regimes as functions of the physical system parameters and
support these rules by extensive numerical calculations. A universal
``phase diagram'' expressed in terms of simple scaling parameters is
shown to be valid for the transition to the regime of full demixing,
inside which we identify several exotic configurations for the two
phase-separated clouds in addition to simple ones consisting of a core
of bosons enveloped by fermions and {\it vice versa}. With increasing
temperature the main role of the growing thermal cloud of bosons is to
transform some exotic configurations into more symmetric ones, until
demixing is ultimately lost. For very high values of boson-fermion
repulsive coupling we also report demixing between the fermions and
the thermally  excited bosons.
\end{abstract}

\section{Introduction}
After the realization of Bose-Einstein condensation (BEC) in atomic
gases \cite{bec}, 
one of the most challenging endeavours in
 experiments on cold atoms is the  cooling of fermionic isotopes of alkali
atoms down to expected
the superfluid transition. Theoretical estimates based on the BCS
model \cite{stoof_BCS} indicate that the
temperature for the superfluid transition should be much lower than
the Fermi temperature $T_f$, which is of the order of the BEC transition
temperature.
In fact, cooling fermions is  harder than cooling bosons: the main
difficulty arises 
from  Fermi statistics, as 
$s$-wave collisions between spin-polarized fermions in a magnetic trap
are forbidden by the Pauli
principle. 
A common strategy uses sympathetic cooling, which is based on  $s$-wave
 collisions  between the fermions and a second
gaseous component made  either of fermions in a different
internal state or of bosons. The latter choice seems to minimize the
effects of Pauli blocking, which limit the process of
cooling  two fermionic components \cite{demarco_PB}. 
Several experiments are currently in progress on trapping and cooling various
boson-fermion mixtures, {\it i.e.} $^6$Li-$^7$Li \cite{hulet_exp,paris_exp},
$^6$Li-$^{23}$Na~\cite{ketterle_BF} and  $^{40}$K-$^{87}$Rb~\cite{jin_BF}.
  The lowest temperature
attained in these experiments so far is  about $0.2\,T_f$.

From the theoretical point of view a mixture of condensed bosons and
 fermions in the normal state is already an interesting system to study, 
because it can show spatial demixing of  the two components
\cite{molmer}. In the homogeneous gas at $T=0$ this is an example of a
 quantum phase transition,
that is a phase transition induced by the interactions \cite{QPT}: on
 increasing the 
boson-fermion repulsion the system minimizes its total energy by
placing the bosons and the fermions (or boson-fermion mixtures of
 different compositions) in different regions of space~\cite{viverit},
even though  this implies a high cost of kinetic energy at the
interface. 
Boson-fermion
demixing in a quasi-spherical trap has been studied 
by Nygaard and M{\o }lmer \cite{molmer_new}, while the conditions for
 demixing inside  a spherical trap  have been set out by two of
 us~\cite{annatosi}  for
$N_b=N_f$
and by Miyakawa {\it 
et al.}~\cite{michenesoio} 
for  $N_f\ll N_b$, with $N_b$ and $N_f$ the numbers of bosons and
 fermions in the trap. The conditions for 
 observing demixing in the Paris
experiment on a $^6$Li-$^7$Li mixture in a cigar-shaped trap~\cite{paris_exp}
has been analyzed by Akdeniz {\it et al.}~\cite{noi}. There are, of
 course, important consequences of finite system size on demixing in
 trapped mixtures, relative to the case of a homogeneous one. In
 particular, the transition is spread out in a finite system and the
 anisotropy of the trap acts differently on the two types of atoms
 (expressions such as phase transition and phase diagram will
 nevertheless continue to be used for brevity in the
 following). Locating the onset of partial demixing is also relevant
 to the practicalities of fermion cooling, since at that point the
 diminishing overlap between the two clouds will start reducing the
 effectiveness of the collisional transfer.

In this paper we give the general conditions under which
phase separation of bosons and fermions occurs inside a harmonic trap at
varying anisotropy and relative numbers of the two components.
Within a mean-field model we make an extensive study of 
the equilibrium density profiles of the bosonic and
fermionic component, ranging from the fully mixed state at small
values of the scattering lengths to the partially demixed state at
intermediate values of the coupling constants and finally to the
regime of full phase separation. We show that the results can be understood
on the basis of scaling laws and simply summarized into
a phase diagram, which is expressed in terms of  scaling parameters
of the system at zero temperature. In the phase-separated regime we
find several configurations having different symmetry and topology, some
of which are metastable, and investigate the role
of the anisotropy of the confinement in determining the
minimum-energy configurations. We then  extend our analysis to
finite temperature, finding that on increasing temperature 
some phase-separated configurations turn or decay into others of higher
symmetry before phase separation disappears. Finally, we show
that  phase separation between the fermions and a bosonic thermal
cloud is also possible in principle on further increase of the
boson-fermion scattering length. 

The paper is organized as follows. In Sec.~\ref{method}
we describe the model that we have used and give the limits
of its applicability, together with  a schematic description of the numerical
method employed to find the phase-separated configurations. 
Sec.~\ref{tzero} summarizes the conditions
under which phase separation occurs at $T=0$,
illustrates various configurations which can be found in
the  phase-separated regime and gives the phase diagram for
the lowest energy configurations.
Sec.~\ref{tfinita} illustrates the effect of temperature on 
the configurations obtained  for both small and large
boson-fermion scattering length. Finally,
Sec.~\ref{secconcl} gives a summary and some concluding remarks.

\section{The method} \label{method} 
We describe the boson-fermion mixture by means of the particle density 
profiles, which are $n_{c}({\bf r})$ for the condensate, $n_{nc}({\bf
r})$ for the bosonic thermal cloud and $n_{f}({\bf r})$ for the
fermions. The components are subject to axially symmetric confining
potentials given by
\begin{equation}
V^{ext}_{b,f}({\bf r})=m_{b,f}\omega^2_{b,f}(r_\perp^2+\lambda^2_{b,f}z^2)/2
\end{equation}
where $m_{b,f}$  are the atomic masses, $\omega_{b,f}$ the trap
frequencies and $\lambda_{b,f}$ the trap
anisotropies. We evaluate the density profiles in a mean-field model
using the Thomas-Fermi approximation for the condensate and the
Hartree-Fock approximation for the other
clouds~\cite{anna_tesi,bf_model}.

The Thomas-Fermi approximation assumes that the number of condensed
bosons is large enough that the kinetic energy term in the
Gross-Pitaevskii equation may be neglected~\cite{baym_pethick}. It yields  
\begin{equation}
n_c({\bf r})=[\mu_b-V_b^{ext}({\bf r})-2gn_{nc}({\bf r})
-fn_f({\bf r})]/g
\label{zehra1}
\end{equation}
for positive values of the function in the square brackets and zero
otherwise. Here, $\mu_b$ is the chemical potential of the bosons and
the coupling constants are $g=4\pi\hbar^2a_{bb}/m_b$ and
$f=2\pi\hbar^2a_{bf}/m_r$, with  $a_{bb}$ and $a_{bf}$  the
boson-boson and boson-fermion
scattering lengths and
$m_r=m_bm_f/(m_b+m_f)$  the reduced mass. The Hartree-Fock
approximation, on the other hand, treats the thermal boson cloud and
the fermion cloud as ideal gases subject to effective potentials
$V_{b,f}^{eff}$, that is 
\begin{equation}
n_{nc,f}({\bf r})=\int\frac{d^3p}{(2\pi\hbar)^3}\left\{\exp\left[\left(
\frac{p^2}{2m_{b,f}}
+V^{eff}_{b,f}({\bf r})-\mu_{b,f}\right)/k_BT\right]\mp 1\right\}^{-1},
\label{zehra4}
\end{equation}
where
\begin{equation}
V^{eff}_{b}({\bf r})=V^{ext}_{b}({\bf r})+2gn_c({\bf r})+2gn_{nc}({\bf r})
+fn_f({\bf r})
\label{zehra2}
\end{equation}
and
\begin{equation}
V^{eff}_{f}({\bf r})=V^{ext}_{f}({\bf r})+fn_c({\bf r})+fn_{nc}({\bf r}).
\label{zehra3}
\end{equation}
The chemical potentials $\mu_{b,f}$ are determined by requiring that
the volume integrals of the densities $n_c({\bf r})+n_{nc}({\bf r})$
and $n_f({\bf r})$ should be equal to the numbers $N_b$ and $N_f$ of
particles.

The model is valid for the bosons when the diluteness condition
$n_ca_{bb}^3\ll 1$ holds and the temperature of the mixture is
sufficiently below the condensation temperature. The fermionic
component has been taken as dilute spin-polarized gas, for which the
fermion-fermion  $s$-wave scattering processes are inhibited  by
the Pauli principle and $p$-wave scattering is
negligible~\cite{pwave}. The condition $k_f a_{bf}\ll 1$ with $k_f$
the Fermi wave number should hold in the mixed regime, but this is not
a constraint in the regime of phase separation where the boson-fermion
overlap is rapidly dropping.

The density profiles at fixed 
numbers $N_b$ and $N_f$ are determined numerically by a
self-consistent solution of the  above equations.
  The profiles are obtained in two
self-consistency loops: first, Eqs.~(\ref{zehra1}) and (\ref{zehra4})
are solved at 
fixed chemical potentials starting from some initial guesses for
the densities; then the chemical potentials are found
iteratively by  a standard algorithm for
multi-variable minimization. Whereas  in the mixed state the 
density profiles at convergence  do not depend on the initial choice,
 different configurations
are found  in the phase-separation regime for the same values of the
parameters by varying the initial 
conditions. We have used this fact  to search for several
possible metastable configurations, which will be illustrated in
Sec.~\ref{total_PS} below. 

\section{Phase diagram at zero temperature}
\label{tzero}

As is evident from the preceding section, a boson-fermion mixture
under confinement is
characterized by  a large number of parameters. 
The aim of this section is to analyze what is the effect of varying
them independently and to provide a unified understanding of the results
using scaling 
laws. 

The interaction energy grows on increasing the values of the
boson-boson and boson-fermion 
coupling constants.
There exists a threshold 
beyond which
the total energy  is minimized by a configuration
in which the two components are spatially separated. 
The transition to the phase-separated regime is
smooth owing to the finite size of the confined gas, and three different
states are indeed recognizable: mixed, partially demixed, and fully
demixed. Depending on the value of the relative strength of the two
coupling constants, it may happen that either the fermions or the
bosons are pushed away from the center of the trap.

\subsection{Partial demixing}
\label{partial_sep}

The boson-fermion interaction energy is proportional to the overlap
between bosonic and fermionic clouds and at $T=0$ we have
\begin{equation}
E_{int}=f \int d^3r \, n_c({\mathbf r}) n_f({\mathbf r}).
\end{equation}
We locate the onset of partial
demixing by looking at the point where the boson-fermion
overlap starts to decrease with increasing $a_{bf}/a_{bb}$.

\begin{figure}
\centerline{
\subfigure[]
{\epsfig{file=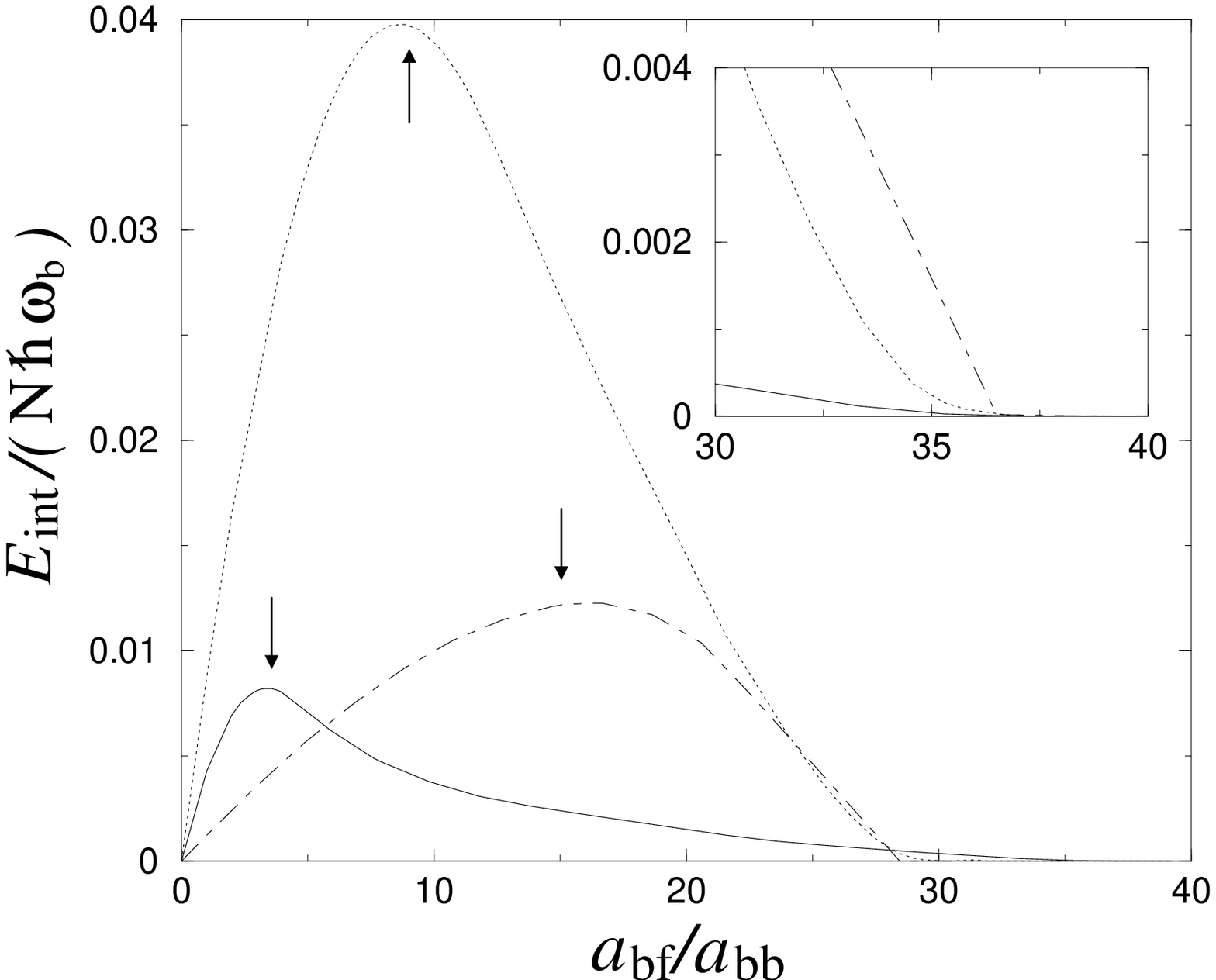,width=0.45\linewidth}\label{eint_a}}
\subfigure[]
{\epsfig{file=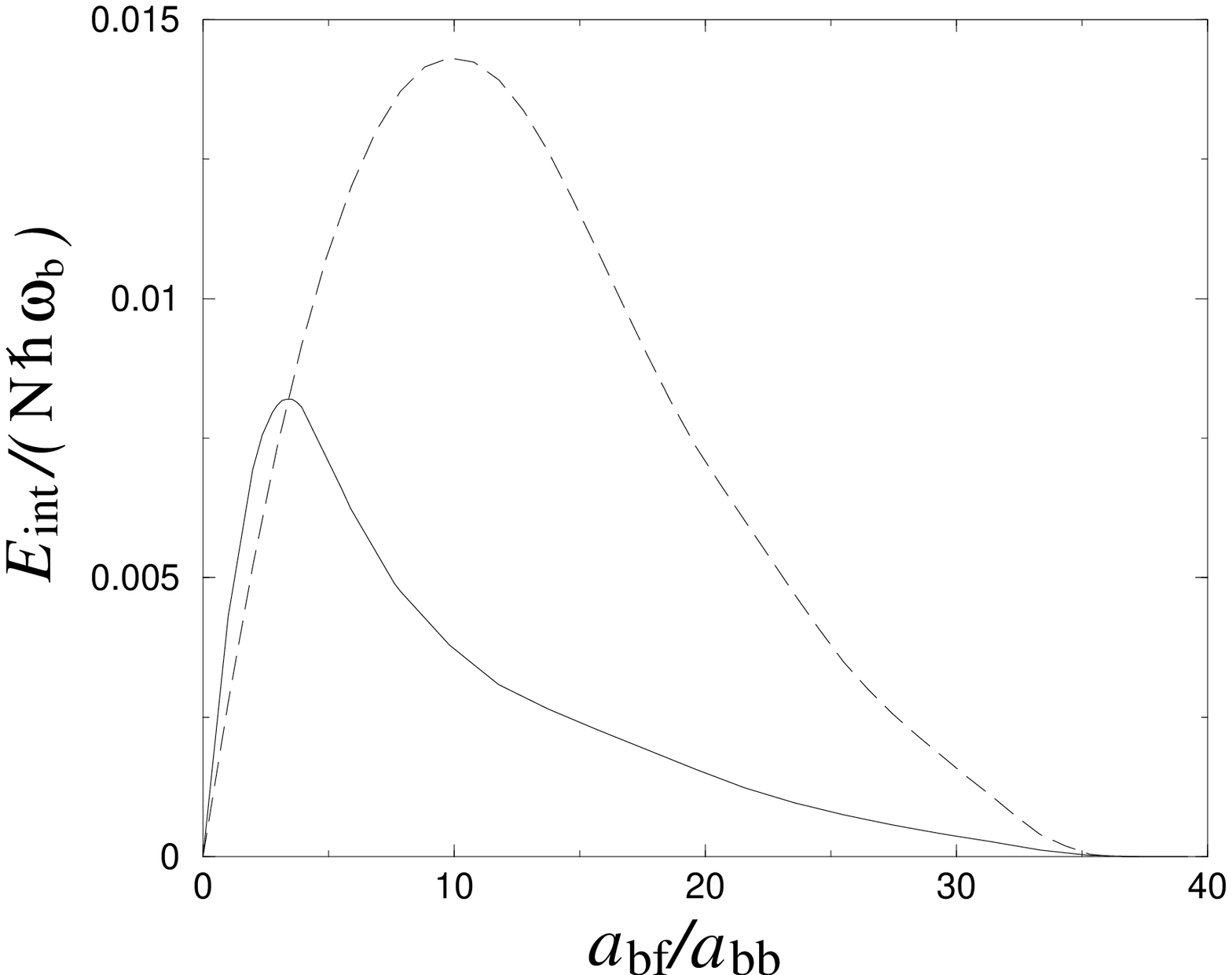,width=0.45\linewidth}\label{eint_b}}}
\caption{Boson-fermion interaction energy at $T=0$ (in units of
$N\hbar\omega_b$ where $N=N_f+N_b$) 
as a function of  the adimensional ratio $a_{bf}/a_{bb}$ for
$a_{bb}=5.1\, a_0$ with $a_0$ being the Bohr radius.
The other parameters are 
$\omega_b/2\pi=4000$~Hz,  
$\omega_f/2\pi=3520$~Hz, $\lambda_b\simeq\lambda_f\simeq 1/60$, and
atomic masses corresponding to the $^6$Li-$^7$Li mixture.
(a) For  
 $N=2.1 \cdot 10^5$, with $N_{f,1}=10^4$ (continuous line),
$N_{f,2}=1.05 \cdot 10^5$ (dotted line) and $N_{f,3}=2\cdot 10^5$ 
(dotted-dashed line). The arrows indicate the positions of the maxima
according to Eq.~(\ref{partially}).
In the inset are shown the same curves in the
same vertical units, 
after rescaling  the second and third curve 
along the horizontal axis by factors $(N_{f,2}/N_{f,1})^{1/12}$ and
$(N_{f,3}/N_{f,1})^{1/12}$.
(b) For fixed number of fermions $N_{f}=10^4$ and
different numbers of bosons
$N_b=2\cdot 10^5$ (continuous line) and $N_b=10^4$ (dashed line).}
\end{figure}

In the following we focus on the simplest case where only one
scattering length is varied, which also  seems realistic from an
experimental point of view. Then the most efficient way to approach
 phase separation is by changing $a_{bf}$ (see Sec.~\ref{total_PS} below).
The behavior of the interaction energy as
a function of the ratio $a_{bf}/a_{bb}$ at fixed $a_{bb}$
is shown in Fig.~\ref{eint_a}. 

The position of the maximum of the interaction energy at fixed
$a_{bb}$ can be estimated
analytically from the condition $\partial E_{int}/\partial
f=0$, by using the approximate expressions
$n_{b,f}\approx N_{b,f}/(4 \pi R_{b,f}^3/3)$ and the
values for the cloud radii $R_f$ and $R_b$ as obtained in the absence
of
 boson-fermion
interactions, $R_f=(48 N_f/\lambda_f)^{1/6}a_{f}$ and $R_b=(15
\lambda_b N_b a_{bb}/a_{b})^{1/5}a_{b}/\lambda_b^{1/3}$ 
with $a_{b,f}=(\hbar/m_{b,f}\omega_{b,f})^{1/2}$. The 
expression for the position of the maximum is 
\begin{equation}
\left.\dfrac{a_{bf}}{a_{bb}}\right|_{max}=\left(c_1\dfrac{N_f^{1/2}}{N_b^{2/5}}+c_2\dfrac{N_b^{2/5}}{N_f^{1/3}}\right)^{-1}
\label{partially}
\end{equation}
where
\begin{equation}
c_1=\dfrac{15^{3/5}}{48^{1/2}}
\dfrac{\lambda_f^{1/2}}{\lambda_b^{2/5}}
\dfrac{m_f^{3/2} \omega_f}{2m_r m_b^{1/2}\omega_b}
\left(\dfrac{a_{bb}}{a_{b}}\right)^{3/5}
\end{equation}
and
\begin{equation}
c_2=\dfrac{48^{1/3}}{15^{3/5}}
\left(\dfrac{6}{\pi}\right)^{2/3}\dfrac{\lambda_b^{2/5}}{\lambda_f^{1/3}}
\dfrac{m_b\omega_b}{2m_r\omega_f}
\left(\dfrac{a_{bb}}{a_{b}}\right)^{2/5}\;.
\end{equation}
We recognise in Eq.~(\ref{partially}) a geometric combination of two
scaling parameters: $c_1{N_f^{1/2}}/{N_b^{2/5}}$
is dominant in the case $N_b\ll N_f$ while
$c_2{N_b^{2/5}}/{N_f^{1/3}}$
was previously identified \cite{lorenzo} in the
regime $N_f\ll N_b$.
The predictions obtained from Eq.~(\ref{partially}) are indicated in
Fig.~\ref{eint_a} by vertical arrows. There clearly is very good
agreement  between the analytical estimate and the numerical results.

\subsection{Full demixing}
\label{total_PS}

The regime of full demixing is taken to be reached when the boson-fermion
overlap becomes negligible. As is shown in Fig.~\ref{eint_b}, within the
three-fluid model the transition point does not depend on the number
of bosons in the trap. This can also  be  predicted from the results
obtained by Viverit {\it et al.}~\cite{viverit} for
the homogeneous mixture, by using  the values of the densities
taken at the center of the trap. Within the same approximation for the
fermion density as in Sec.~\ref{partial_sep} we obtain the condition
for phase separation at $T=0$ as
\begin{equation}
 \alpha \, k_fa_{bb}>\left(\frac{a_{bb}}{a_{bf}}\right)^2,
\label{condition}
\end{equation}
where $k_f=(48 \lambda_f N_f)^{1/6}/a_{f}$ and $\alpha=
[3^{1/3}/(2\pi)^{2/3}] m_b m_f/(4 m_r^2)$. Equation~(\ref{condition})
is valid for different numbers of bosons and fermions as well as for
different atomic masses and trap frequencies, and agrees very well
with the phase-separation criterion given earlier by two of
us~\cite{annatosi} for the case where $m_f=m_b$ and
$\omega_f=\omega_b$. It confirms that 
the two relevant parameters for describing the
transition to the fully demixed  regime are $a_{bf}/a_{bb}$ and
$\alpha k_f a_{bb}$.
We also immediately find that the value of $a_{bf}$ at the
transition point scales as $N_f^{-1/12}$ at fixed $a_{bb}$:
this result is illustrated in the inset in  Fig.~\ref{eint_a}. 

\begin{figure}
\centerline{
\epsfig{file=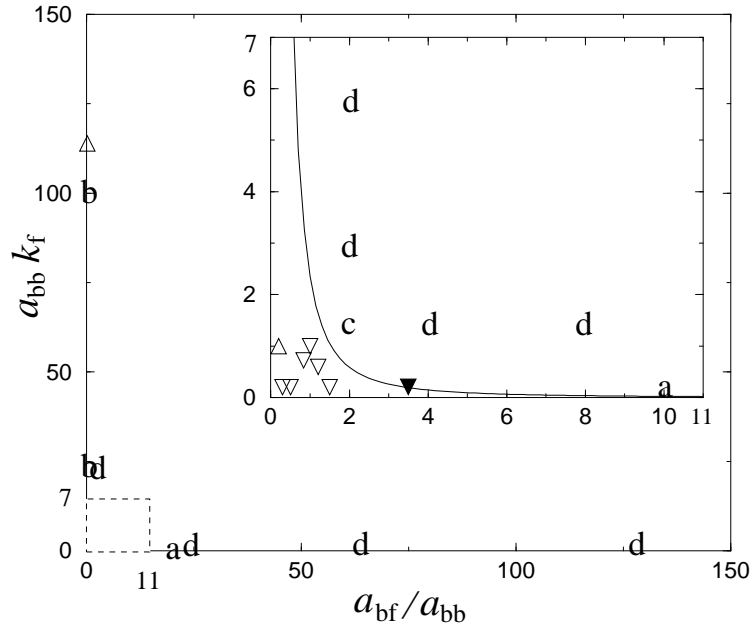,width=0.55\linewidth}}
\vspace{0.2cm}
\caption{Phase diagram at $T=0$, in the plane defined
by the adimensional parameters $a_{bb}k_f$ and $a_{bf}/a_{bb}$.
The bottom left corner of the figure is shown enlarged in the inset.
The continuous line in the inset corresponds to the condition of 
phase separation 
given by Eq.~(\ref{condition}). In the mixed-phase regime we have the
following configurations:
 $\vartriangle$  fermions at the center:
$\triangledown$  bosons at the
center; $\blacktriangledown$ bosons an the center in an almost demixed
configuration. For the 
phase-separated regime we have used the following notations (see 
Fig.~\ref{fig_zoo}):
(a) symmetric configuration with fermions outside;
(b) symmetric configuration with bosons outside;
(c) boson torus inside a fermion cloud;
(d) three-component
symmetric configuration with fermions outside and in
the central core. All the configurations in this phase diagram have
been evaluated at zero temperature, using $N_b=N_f=10^4$,
$\omega_b/2\pi=4000$~Hz,  
$\omega_f/2\pi=3520$~Hz, $\lambda_b\simeq\lambda_f\simeq 1/60$,  and
atomic masses corresponding to the $^6$Li-$^7$Li mixture.
} 
\label{fig3}
\end{figure}

Equation~(\ref{condition}) has been verified by the numerical solution
of the three-fluid model for a variety of  different sets of parameters. 
First of all we have studied the location of  phase separation in
parameter space by varying the
scattering lengths over much wider ranges than allowed by 
the usual off-resonant values ({\it e.g.}
in the $^6$Li-$^7$Li
mixture),  as might be attained experimentally by exploiting
optically or magnetically induced Feschbach
resonances \cite{feschbach}.
For these calculations we have chosen the geometrical parameters of the
traps and the numbers of atoms as in the 
Paris experiment on $^6$Li-$^7$Li~\cite{paris_exp}: namely $N_b \simeq
N_f=10^4$, 
$\omega_b/2\pi=4000$~Hz, 
$\omega_f/2\pi=3520$~Hz and $\lambda_b\simeq\lambda_f\simeq 1/60$. 
The results are summarized  in a quantum  phase diagram at $T=0$ in the plane 
$\{a_{bf}/a_{bb},\, k_f a_{bb}\}$, which is reported in
Fig.~\ref{fig3}. This contains 
all the lowest-energy configurations that we have found, with the
inset showing an enlargement of the region near the origin of the
above-mentioned plane. The curve drawn inside the inset in
Fig.~\ref{fig3} is obtained from Eq.~(\ref{condition}) and is in full
agreement with the numerical results in separating the region of full
demixing from the mixing or partial-demixing regions.

\begin{figure}
\centerline{
\epsfig{file=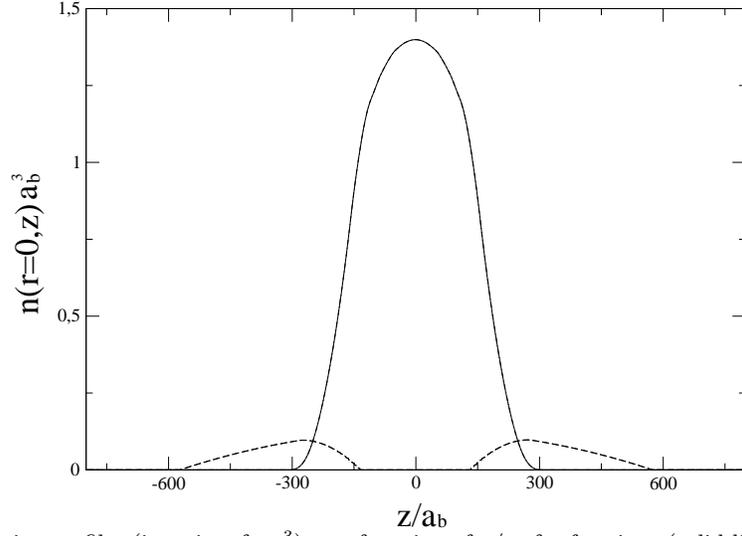,width=0.55\linewidth}}
\caption{Section of the density profiles  (in
units of $a_{b}^{-3}$)  as a function of $z/a_{b}$ for 
fermions (solid line) and bosons (dashed line) in a configuration
where  a mixed phase and  a purely bosonic one coexist
($\vartriangle$ in Fig.~\ref{fig3}).
This configuration has been obtained at $T=0$ with  $a_{bb}=323000\,a_0$
and $a_{bf}=32300\,a_0$. The numbers of atoms and all other parameters are
as in Fig.~\ref{fig3}.}
\label{fig4}
\end{figure}

We turn to a detailed account of the configurations corresponding to
the various symbols and letters in the phase diagram in
Fig.~\ref{fig3}. Below
 the phase separation 
line in the inset  two different types of 
configurations are found, one  
with a core of bosons in partial overlap with  an envelope of mainly
fermions (triangles down) and a specular
one with a core of fermions  enveloped by mainly bosons (triangle
up). 
An interesting
configuration that we have observed in the latter case 
shows coexistence of a mixed state with  a purely bosonic
phase (see Fig.~\ref{fig4}): this combination is not 
stable in the homogeneous mixture~\cite{viverit}.

\begin{figure}
\centerline{
\epsfig{file=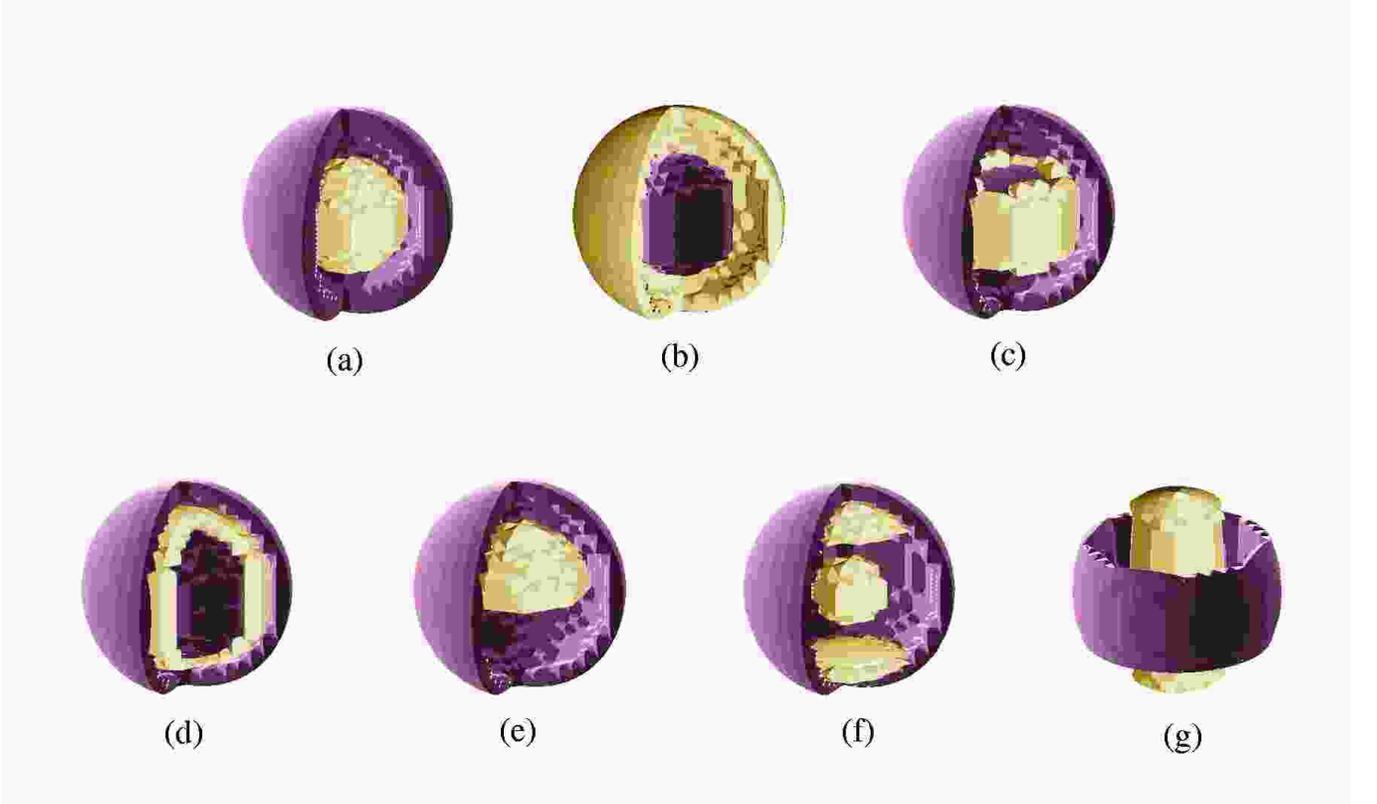,width=1\linewidth}}
\caption{Configurations in the phase-separated regime for $N_f=
N_b=10^4$ particles at $T=0$.
(a) symmetric
configuration with fermions outside; (b) symmetric
configuration with bosons outside; (c) boson torus inside fermion 
cloud; (d)
threefold symmetric configuration; (e)
asymmetric configuration; (f)  boson sandwich
inside a fermion cloud; and (g) torus of fermions around an elongated  core
of bosons. The structures (a) and (c-f) are for the same 
values of $a_{bb}=600\,a_0$ and  $a_{bf}=6000\,a_0$, and their 
energies per particle are 4.24, 4.71, 5.38, 5.50, and  5.85 $\hbar \omega_b$.
The configuration (b) was obtained for $a_{bb}=64600\,a_0$ and 
$a_{bf}=32300\,a_0$ and the value of its energy per particle is 13.25
$\hbar \omega_b$.
The configuration (g) was obtained for $a_{bb}=6250\,a_0$ and 
$a_{bf}=12500\,a_0$ 
and the value of its energy per particle is 9.27 $\hbar \omega_b$.  All  
other parameters are  as in Fig.~\ref{fig3}. 
} 
\label{fig_zoo}
\end{figure}

In the regime of phase separation a variety of 
different configurations are observed, including several that require
a  break of the
symmetry imposed by  the confining harmonic potentials.
The letters a-d in the phase diagram in Fig.~\ref{fig3} give the
locations of four such energetically stable structures, which are
shown in Fig.~\ref{fig_zoo} together with some other metastable
structures. The symmetric configuration with boson inside and fermion
outside, indicated by (a), is obtained in  Fig.~\ref{fig_zoo} with the
choice $a_{bb}=600$ Bohr radii and $a_{bf}=10\, a_{bb}$, while the
complementary configuration (b) formed from fermions inside and bosons
outside is obtained with  $a_{bb}=64600$ Bohr radii and
$a_{bf}=a_{bb}/2$. Other configurations obtained with the same
scattering lengths as the symmetric structure of type (a) are shown in
Fig.~\ref{fig_zoo}: these are (c) a boson torus inside a fermion
cloud, (d) a threefold symmetric structure formed from fermions
surrounded by a shell of bosons inside a fermion envelope, (e) an
asymmetric structure in which the bosons are shifted away from the
center of the trap along the $z$ axis, and (f) a ``sandwich''
formed by bosons inside a fermion cloud. 
Finally the structure
indicated by (g) in Fig.~\ref{fig_zoo}, which consists of a torus of
fermions around an elongated boson core, is found with the choice
$a_{bb}=6250$ Bohr radii and $a_{bf}=2 \, a_{bb}$. 

\begin{figure}
\centerline{
\epsfig{file=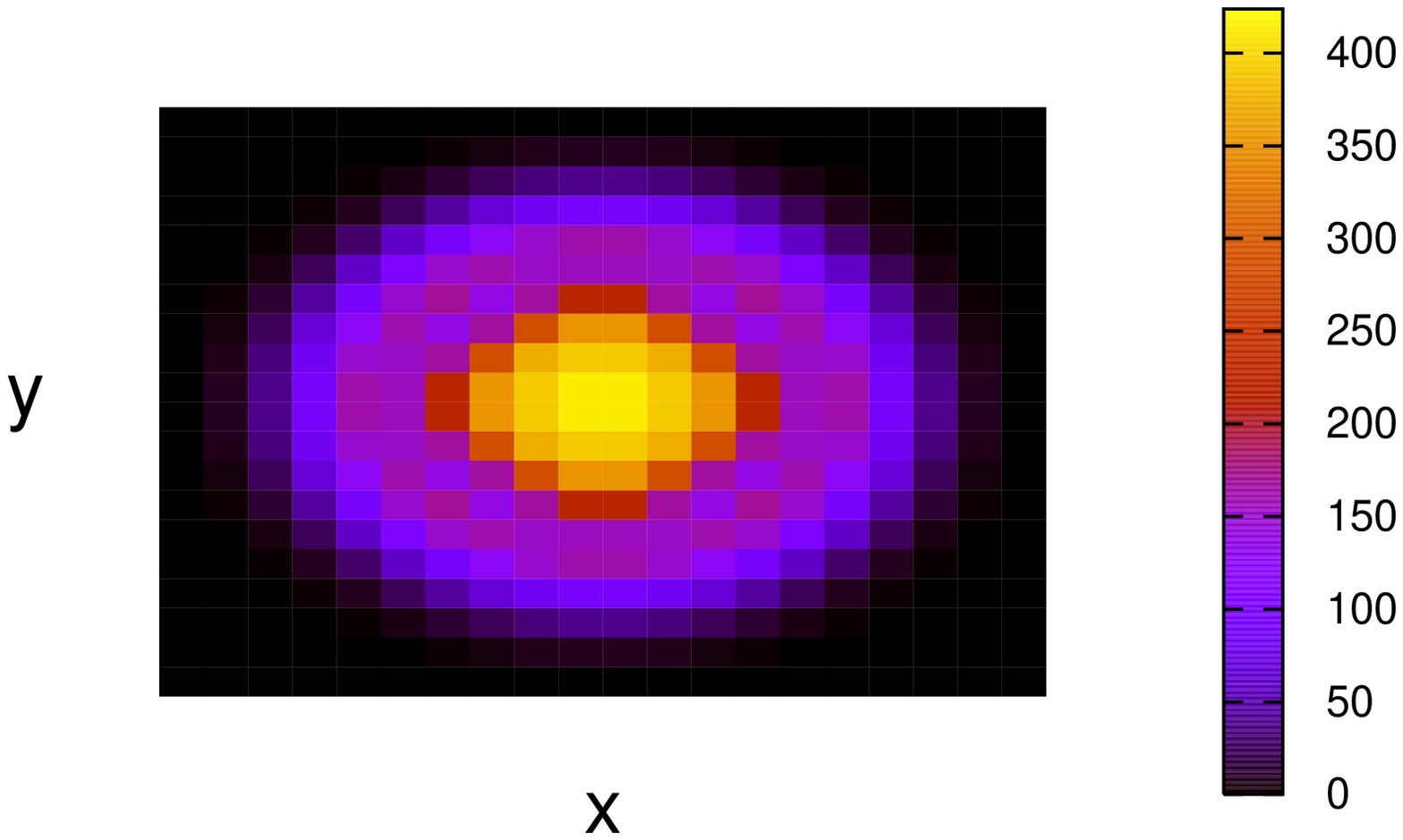,width=0.4\linewidth,angle=270}
\epsfig{file=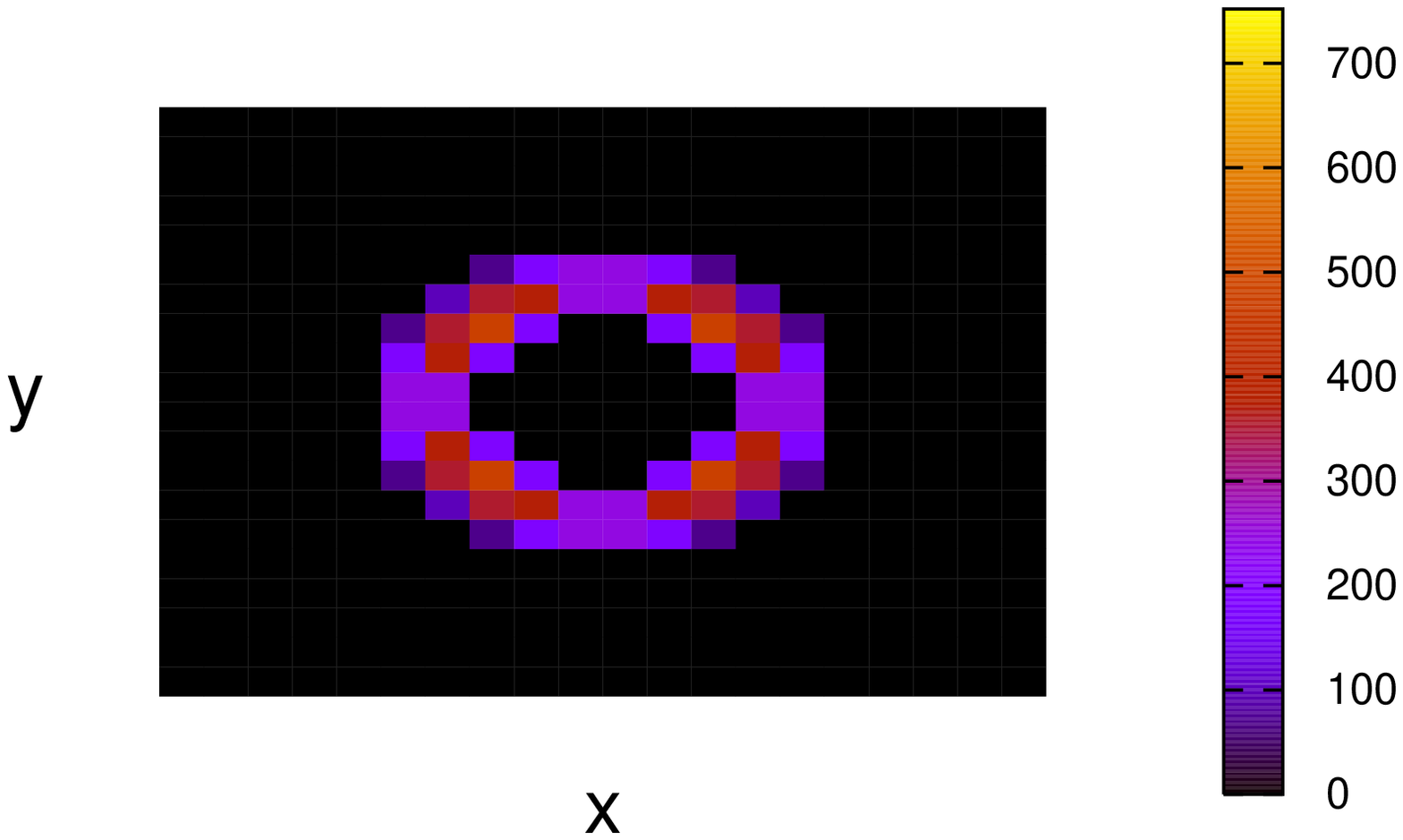,width=0.4\linewidth,angle=270}}
\vspace{-2cm}
\centerline{
\epsfig{file=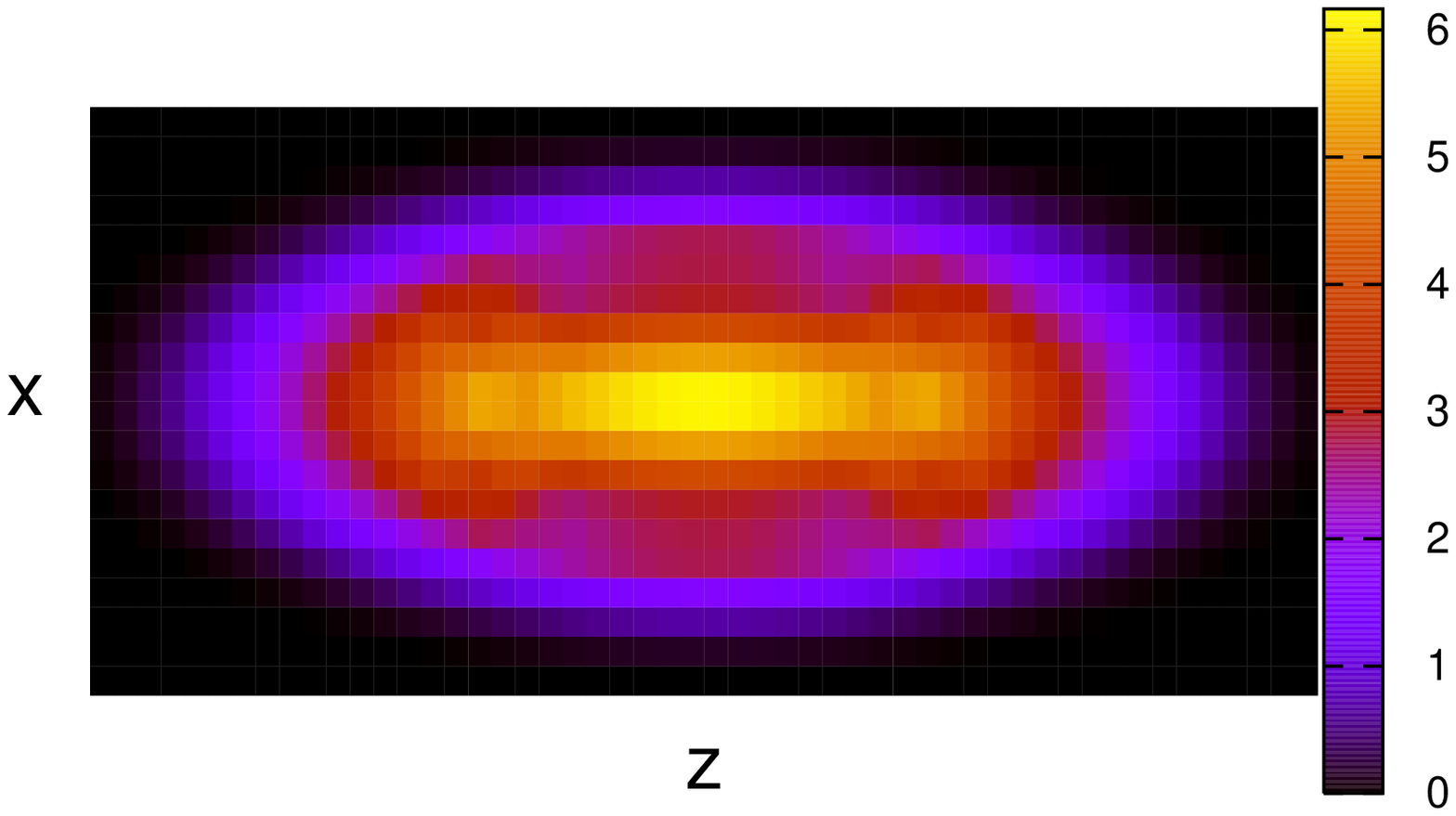,width=0.4\linewidth,angle=270}
\epsfig{file=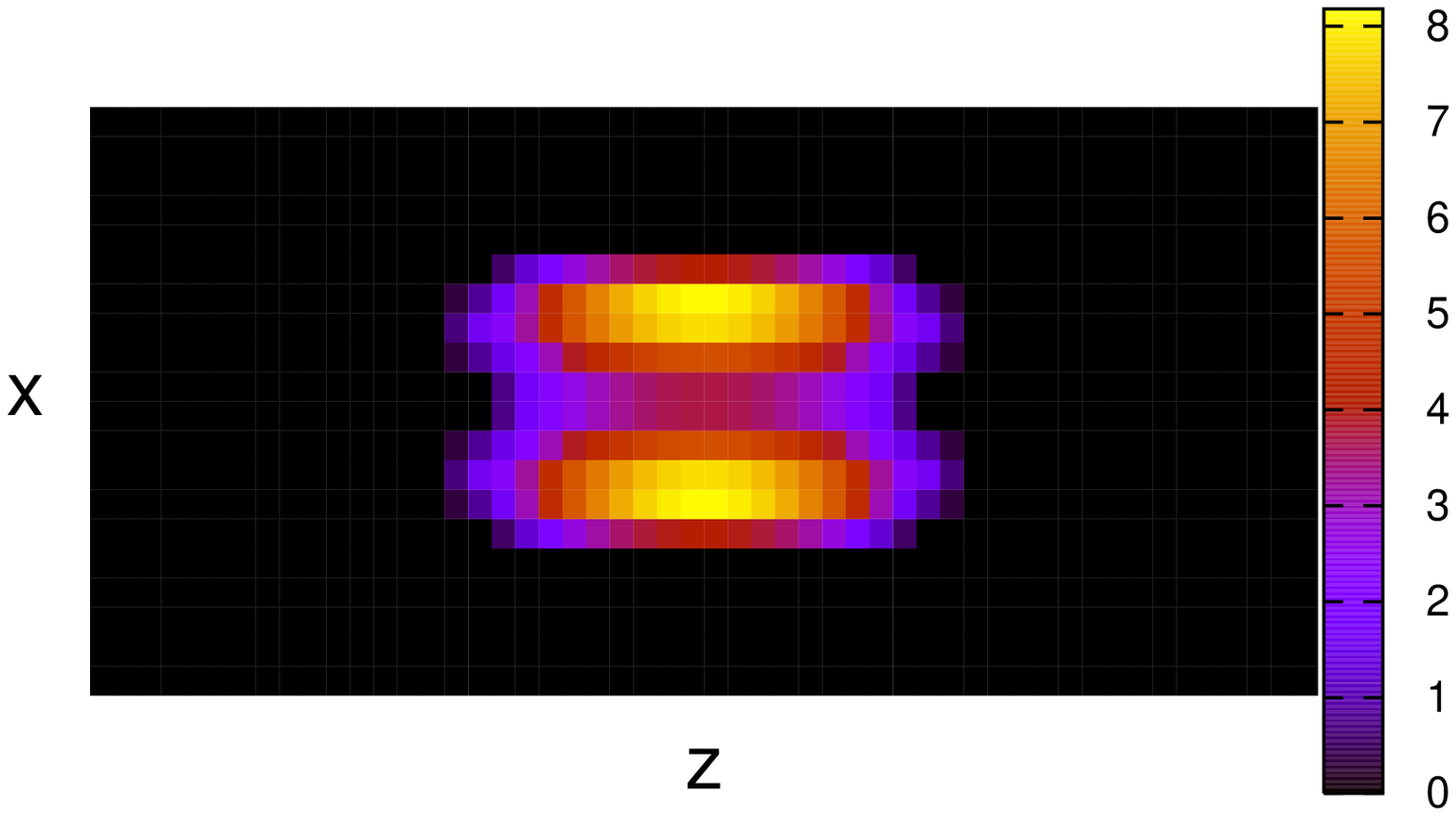,width=0.4\linewidth,angle=270}}
\vspace{-1cm}
\caption{Column densities  (in units of $a_{b}^{-2}$) 
in the radial plane (top) and in the   $\{x,z\}$ plane (bottom)
for  
the fermionic cloud (left) and the bosonic cloud (right)
of the bosonic   torus  (configuration (c) in Fig.~\ref{fig_zoo},
obtained with
$a_{bb}=600\,a_0$ and  $a_{bf}=6000\,a_0$).
The size of the figures in the $\{x,y\}$ plane and 
in the $\{x,z\}$ plane
is $11\mu m\times11\mu m$ and $11\mu m\times 400\mu m$,  respectively.
The other parameters are as in Fig.~\ref{fig3}.}
\label{fig2a}
\end{figure}

\begin{figure}
\centerline{
\epsfig{file=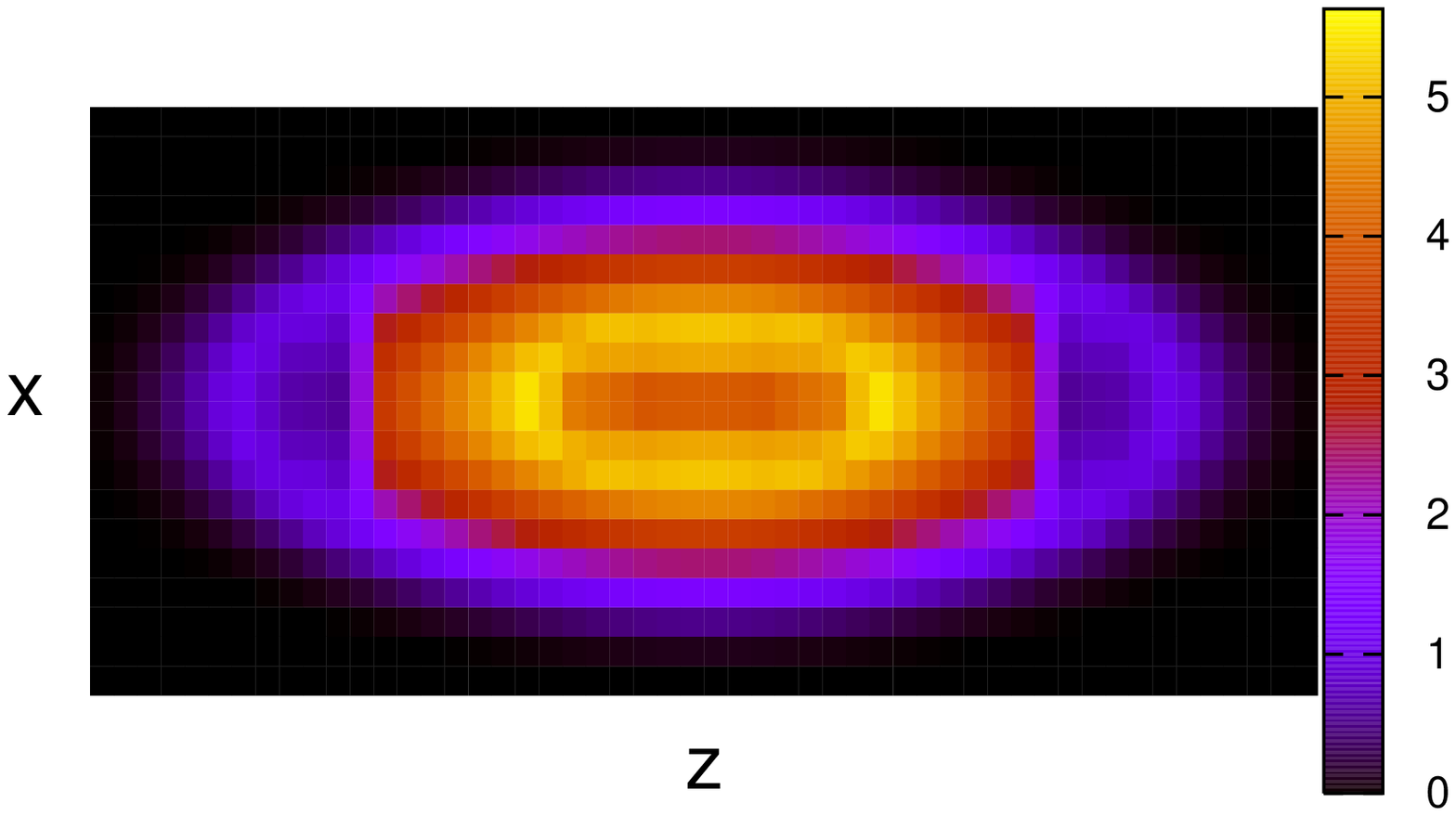,width=0.4\linewidth,angle=270}
\epsfig{file=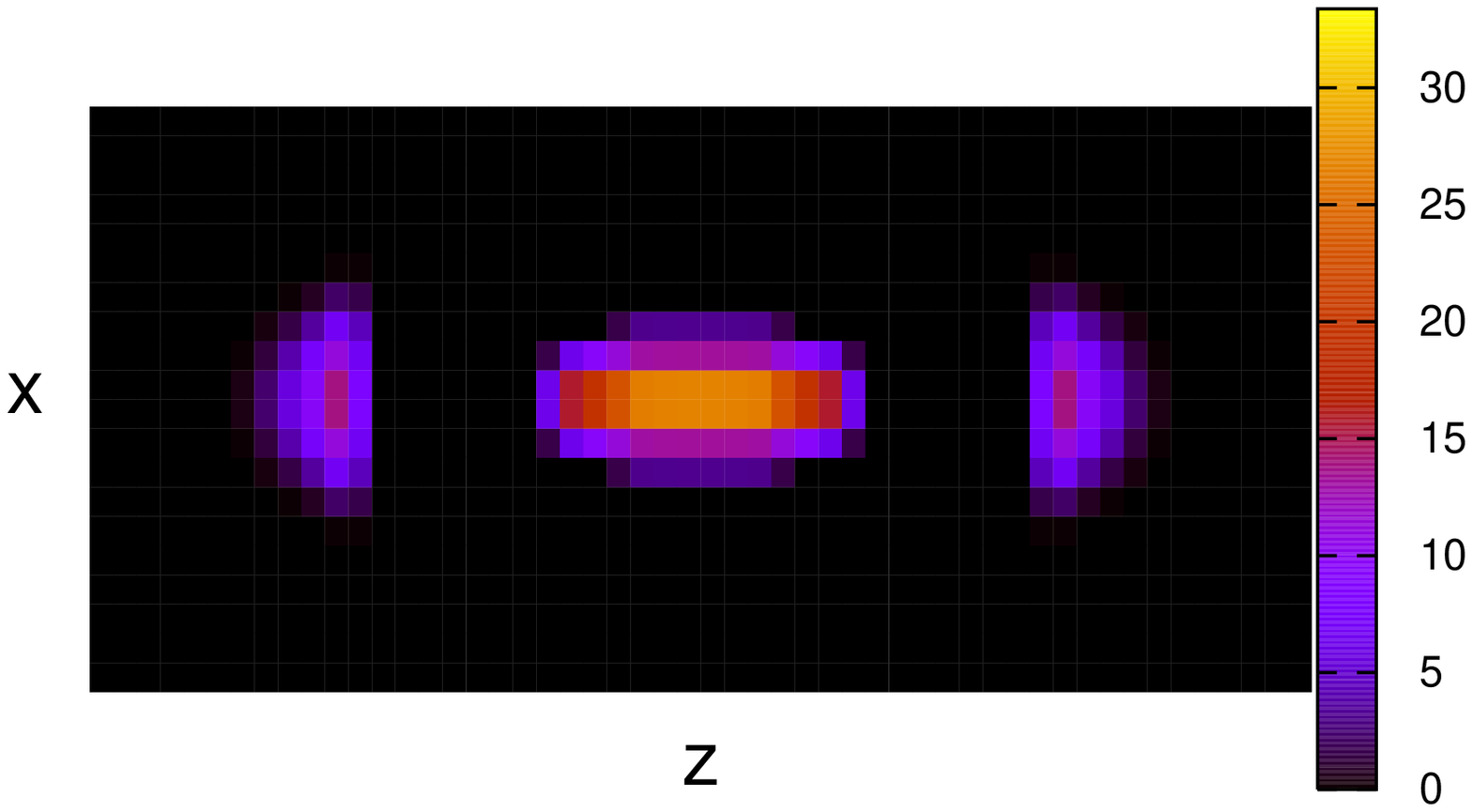,width=0.4\linewidth,angle=270}}
\vspace{-1cm}
\caption{
Column densities  in the $\{x,z\}$ plane (in units of $a_{b}^{-2}$)
for  the fermions 
(left) and for the condensate (right) in 
the sandwich configuration (f) in Fig.~\ref{fig_zoo}, obtained with
$a_{bb}=600\,a_0$ and  $a_{bf}=6000\,a_0$.
The size of the figures 
is $11\mu m\times 400\mu m$. The other parameters are as in
Fig.~\ref{fig3}. 
}
\label{fig2}
\end{figure}

With the exception of (a) and (b) the configurations shown in
Fig.~\ref{fig_zoo}
are very
different from the density distributions that are met in the mixed phase,
and thus could be  identified in an experiment by looking at 
the column-density pictures of the atomic clouds. As an illustration we give
in Fig.~\ref{fig2a} the column-density images corresponding to the
case (c) of a  bosonic torus, and in 
Fig.~\ref{fig2} those for the 
case (f) of a boson sandwich.

Finally, we have examined the role of the anisotropy of the confinement
on 
the configurations of lowest energy in the phase-separated regime.  
To this end we have evaluated the stability of the lowest-energy
configuration by varying the anisotropy
from $\lambda_f=1/60$ to $\lambda_f=1$ while keeping the ratio
$\lambda_b/\lambda_f\simeq 1$, with the choice 
$a_{bb}=600$ Bohr radii and $a_{bf}=10\, a_{bb}$ for the scattering
lengths.
The  most stable configuration remains the symmetric structure (a) except 
that for $\lambda_f\simeq 1$ the structure  (d) has the lowest 
energy.
For this particular choice of parameters the
relative difference in energy between the metastable configurations 
does not vary much with $\lambda_f$.

\section{The role of temperature}
\label{tfinita}
At finite temperature an increasing fraction of the bosons populates
the thermal cloud, thus depleting the condensate. In harmonic
confinement the thermal cloud is wider and more dilute than the
condensate, and thus is much less interacting with the fermions. As
a result the thermal cloud is
not yet in phase separation when the condensate already is.

\begin{figure}
\centerline{
\epsfig{file=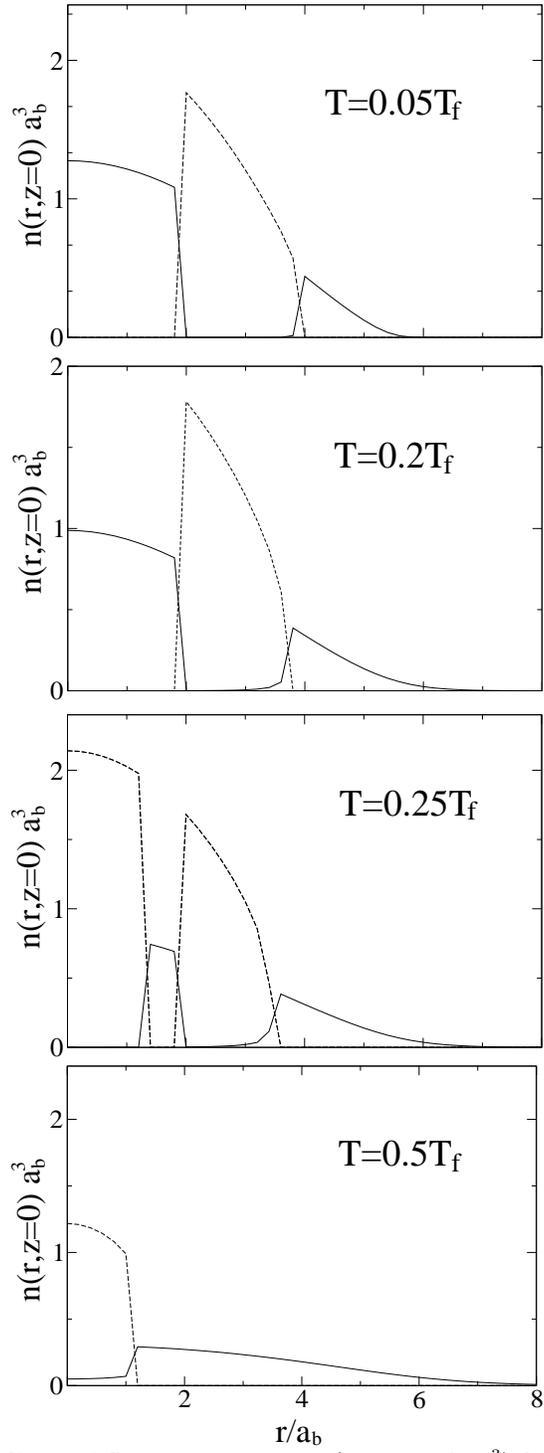,width=0.4\linewidth}}
\caption{Sections of the density profiles at different temperatures
(in units of 
 $a_{b}^{-3}$)
for the fermions (solid line) and for the 
condensate (dashed line) as functions of the radial coordinate
 $r/a_{b}$
 for the configuration (c) in
 Fig.~\ref{fig_zoo}, obtained with  $a_{bb}=600\,a_0$ and  $a_{bf}=6000\,a_0$. 
The values of the other parameters are as in Fig.~\ref{fig3}.
}
\label{fig5}
\end{figure}

At low temperatures the non-condensate fraction is small and its
presence does not
affect the description of the phase separation between the condensate and
the fermions that we have given in the preceding section. In particular we have
verified that at $T$ of the order of $0.2\, T_f$, as are
reached in  current experiments, all the exotic configurations
(c) to (g) remain possible. They  are not
found at higher temperatures: we show as an example in Fig.~\ref{fig5}
the behavior of the boson torus configuration (c)  as a function of
temperature.  For $0.2\,T_f<T\le
0.3\,T_f$ a part of the condensed bosons  moves from the torus to 
 the center of the trap
and  
for $T>0.3T_f$ the phase separation is lost.

\begin{figure}
\centerline{
\epsfig{file=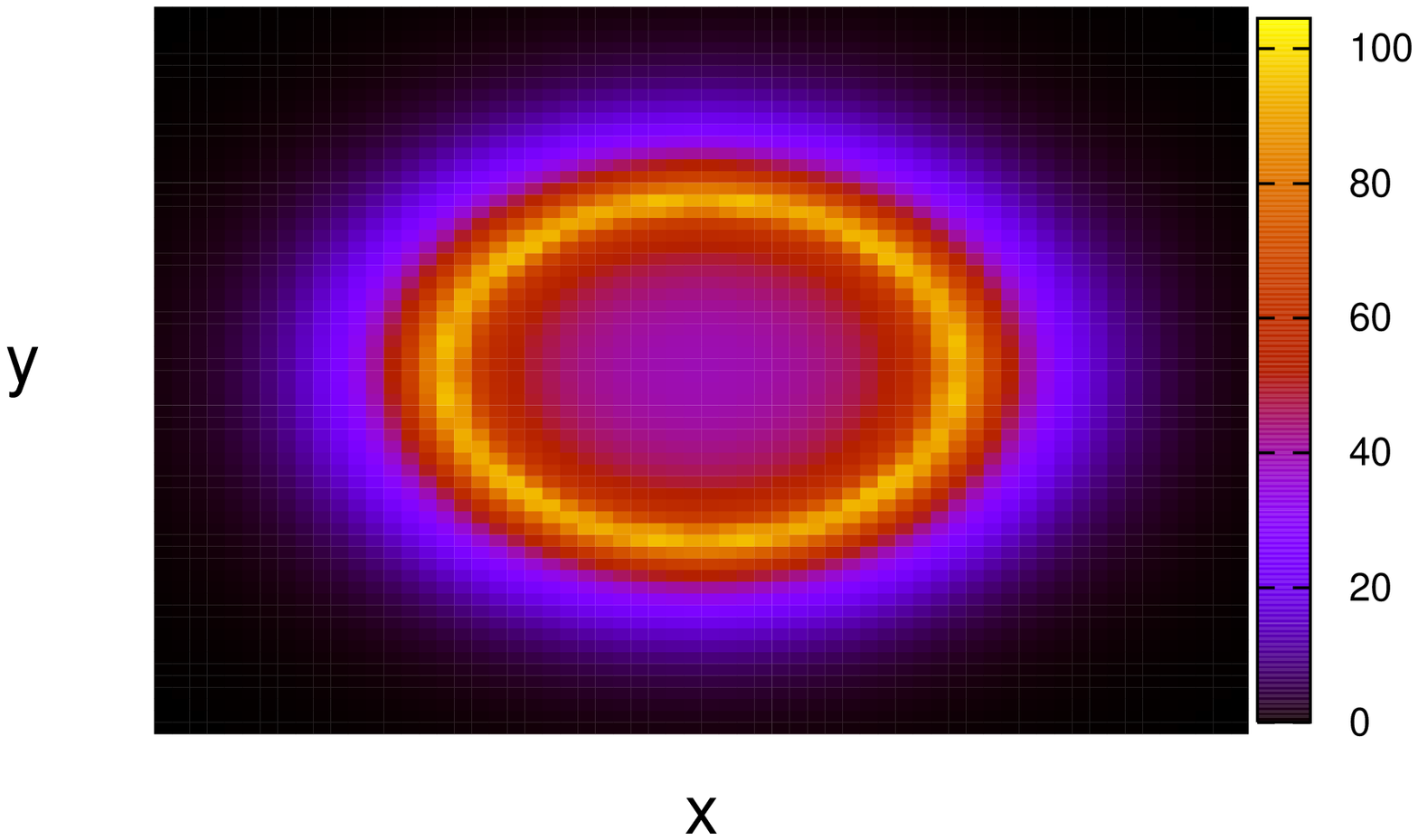,width=0.4\linewidth,angle=270}
\epsfig{file=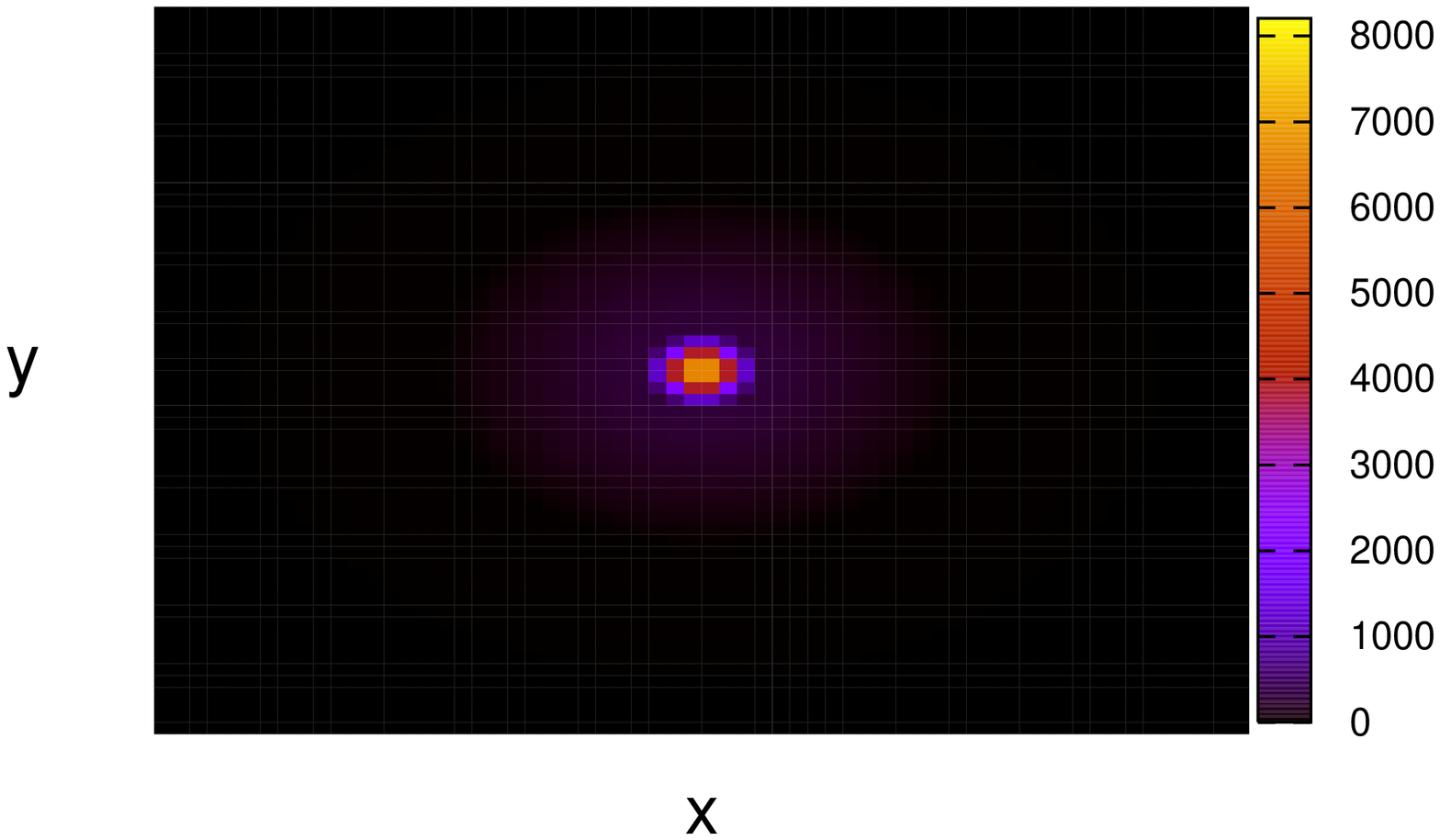,width=0.4\linewidth,angle=270}}
\caption{Column densities  in the
$\{x,y\}$ plane (in units of $a_{b}^{-2}$)
for  fermions (left) and  bosons (right)  
in the fully demixed 
 regime at $T=0.6\,T_f$,
for $a_{bb}=5.1\,a_0$ and $a_{bf}=2\cdot 10^5\,a_0$. The values of
the other parameters are as in Fig.~\ref{fig3}.
In the right panel the thermal bosons are not visible because of 
the high density of the central condensate.
The size of the figures
is $11\mu m\times11\mu m$.
}
\label{fig_last}
\end{figure}

At still  higher temperatures the thermal cloud starts to be consistently
populated and  we have found that phase separation between the fermions
and the entire bosonic cloud becomes possible for very large values of
$a_{bf}/a_{bb}$ \cite{noi}. In this case, as is illustrated in
Fig.~\ref{fig_last}, 
the fermion cloud exhibits a central hole having the size of the
bosonic thermal cloud.

\section{Summary and concluding remarks}
\label{secconcl}

We have performed an extensive analysis of the phenomenon of
phase separation for a boson-fermion mixture under confinement 
on the basis of the atomic density profiles of the two components.
First of all we have pointed out that  three regimes can be identified
owing to the finite size of the
system: a mixed phase where the
bosonic and fermionic clouds overlap, a partially demixed state where
the overlap is decreasing, and a phase-separated
regime where there is no overlap between the two components at zero
temperature. 
We have given analytical expressions for the positions of the two
cross-overs between these regimes 
in terms of the physical parameters of the system, and
verified these conditions by an extensive numerical study based on
varying the coupling constants, the numbers of atoms, and the geometry of the
trap. General expressions for the transition lines are particularly
useful in view of the several experiments in progress
\cite{paris_exp,ketterle_BF,jin_BF}, which are run with different
experimental parameters such as the 
relative numbers of bosons and fermions, the atomic masses, and the trap
anisotropies. 

The transition to the partially phase-separated regime,
that we have located where the 
boson-fermion interaction energy starts to decrease on increasing the
boson-fermion scattering, is important for the
experiments on fermion cooling:
the boson-fermion collision rate is decreased by a geometrical
factor which is proportional to the overlap between the two clouds.
At fixed boson-boson scattering length we have
given a simple model for the transition in terms of a
combination of two scaling parameters, which characterize 
the cases $N_b \ll N_f$ and $N_f \ll N_b$.

The transition to full demixing has been illustrated by a
universal phase diagram expressed in terms of  scaling parameters
of the system. It is remarkable that 
in our model the location of this transition  does not depend on
the number of  bosons. 
In the region of phase separation we have found several configurations 
with various topology  and identified the minimum-energy
configurations. 
We have also investigated the effect of increasing temperature on  phase
separation. Some exotic
configurations that are found in the 
phase diagram at $T=0$
turn into others of higher symmetry before the condensate-fermion
demixing is lost.

The main limitations of the present model come from  the two basic
approximations that we have employed: the Thomas-Fermi approximation
for the condensate and the mean-field Hartree-Fock
approximation for the other gaseous clouds.  The Thomas-Fermi
approximation breaks down for a 
very small
condensate or for negative values of the boson-boson scattering length, and 
one would need to solve the full Gross-Pitaevskii equation. A
description   beyond mean field is required when
the dilution parameters increase and 
for a trapped gas this is still
an open problem.  The analogue of the
Beliaev expansion at $T=0$ has  recently been derived for the homogeneous 
boson-fermion mixture \cite{fabrizio}.

Of course, the evolution of the mixture towards phase separation
modifies not only its equilibrium properties, but also its dynamical
properties and collective
excitation spectrum.  The approach to
the transition will be signalled by a softening of the frequencies
 of  surface or 
bulk modes having the appropriate symmetry for a
phase-separated configuration to be attained. Experiments
measuring dynamical properties
could therefore provide an alternative method for
revealing the onset of demixing. Calculations on these spectra are in
progress.

\acknowledgements
This work was supported by INFM through the PRA2001-Photonmatter.

\newpage

\end{document}